\documentclass[
 reprint,
 amsmath,amssymb,
 aps,
 prl
]{revtex4-1}

\usepackage{graphicx}
\usepackage{dcolumn}
\usepackage{bm}

\newcommand{\bra}[1]{\langle #1|}
\newcommand{\ket}[1]{|#1\rangle}

\begin{document}

\title{Violation of Heisenberg's Measurement-Disturbance Relationship by Weak Measurements}
 
\author{Lee A. Rozema}

\author{Ardavan Darabi}
 
\author{Dylan H. Mahler}

\author{Alex Hayat}

\author{Yasaman Soudagar}
 
\author{Aephraim M. Steinberg}

\affiliation{%
Centre for Quantum Information \& Quantum Control and Institute for Optical Sciences,
Dept. of Physics, 60 St. George St., University of Toronto, Toronto, Ontario, Canada M5S 1A7
}

\date{\today}
\begin{abstract}
While there is a rigorously proven relationship about uncertainties intrinsic to any quantum system, often referred to as ``Heisenberg's Uncertainty Principle,'' Heisenberg originally formulated his ideas in terms of a relationship between the precision of a {\it measurement} and the disturbance it must create. Although this latter relationship is not rigorously proven, it is commonly believed (and taught) as an aspect of the broader uncertainty principle. Here, we experimentally observe a violation of Heisenberg's ``measurement-disturbance relationship", using weak measurements to characterize a quantum system before and after it interacts with a measurement apparatus. Our experiment implements a 2010 proposal of Lund and Wiseman to confirm a revised measurement-disturbance relationship derived by Ozawa in 2003. Its results have broad implications for the foundations of quantum mechanics and for practical issues in quantum mechanics.
\end{abstract}

\maketitle

The Heisenberg Uncertainty Principle is one of the cornerstones of quantum mechanics. In his original paper on the subject, Heisenberg wrote ``At the instant of time when the position is determined, that is, at the instant when the photon is scattered by the electron, the electron undergoes a discontinuous change in momentum. This change is the greater the smaller the wavelength of the light employed, i.e., the more exact the determination of the position'' \cite{1}. Here Heisenberg was following Einstein's example and attempting to base a new physical theory only on observable quantities, that is, on the results of measurements. The modern version of the uncertainty principle proved in our textbooks today, however, deals not with the precision of a measurement and the disturbance it introduces, but with the {\it intrinsic} uncertainty any quantum state must possess, regardless of what measurement (if any) is performed \cite{2,3,4}. These two readings of the uncertainty principle are typically taught side-by-side, although only the modern one is given rigorous proof. It has been shown that the original formulation is not only less general than the modern one -- it is in fact mathematically incorrect \cite{5}. Recently, Ozawa proved a revised, universally valid, relationship between precision and disturbance \cite{6}, which was indirectly validated in \cite{7}. Here, using tools developed for linear-optical quantum computing to implement  a proposal due to Lund and Wiseman \cite{23}, we provide the first direct experimental characterization of the precision and disturbance arising from a measurement, violating Heisenberg's original relationship.
\begin{figure}
\includegraphics[scale=0.65]{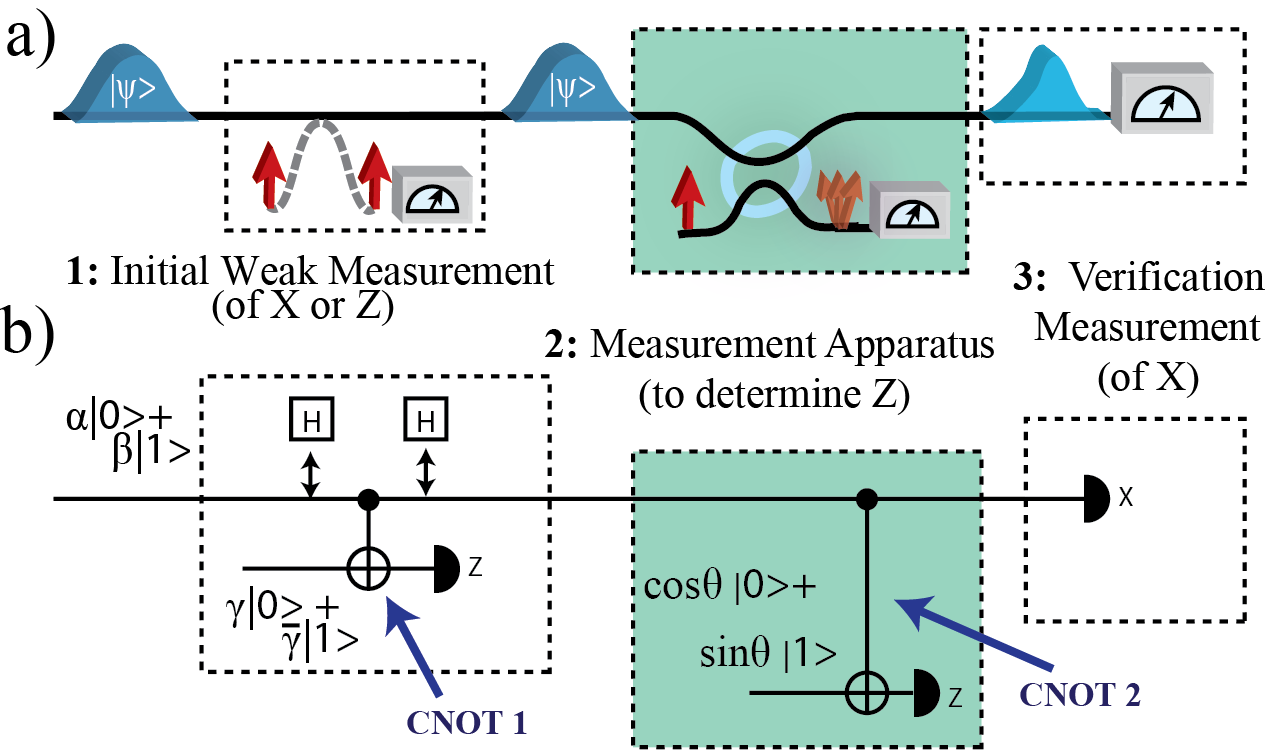}
\caption{Schematic of the weak measurement proposal. a) A general method for measuring the precision and disturbance of any system. The system is weakly measured before the measurement apparatus and then strongly measured afterwards. b) A quantum circuit, proposed by Lund and Wiseman in \cite{23}, which can be used to measure the precision and disturbance of $\hat{X}$ and $\hat{Z}$ for a qubit system. We can weakly measure $\hat{X}$, rather than $\hat{Z}$, by inserting Hadamard gates before and after the control.}
\end{figure}

In general, measuring one observable (such as position, q) will, according to quantum mechanics, induce a random disturbance in the complementary observable (in this case momentum, p). Heisenberg proposed, and it is widely believed, that the product of the measurement precision, $\epsilon(q)$, and the magnitude of the induced disturbance, $\eta(p)$, must satisfy $\epsilon(q)\eta(p)\approx h$, where $h$ is Planck's constant.  This idea was at the crux of the Bohr-Einstein debate \cite{8}, and the role of momentum disturbance in destroying interference has remained a subject of heated discussion \cite{9,10,11}.  Recently, the study of uncertainty relations in general has been a topic of growing interest, specifically in the setting of quantum information and quantum cryptography, where it is fundamental to the security of certain protocols \cite{12,13}.  The relationship commonly referred to as the Heisenberg Uncertainty Principle (HUP) - in fact proved later by Weyl \cite{4}, Kennard \cite{3}, and Robertson \cite{2} - refers not to the precision and disturbance of a measurement, but to the uncertainties intrinsic in the quantum state. The latter can be quantified by the standard deviation  $\Delta \hat{A} = \sqrt{\ket{\psi}\hat{A}^2\bra{\psi} -\ket{\psi}\hat{A}\bra{\psi}^2}$, which is independent of any specific measurement. This relationship, generalized for arbitrary observables $\hat{A}$ and $\hat{B}$, reads:
\begin{figure}
\includegraphics[scale=0.63]{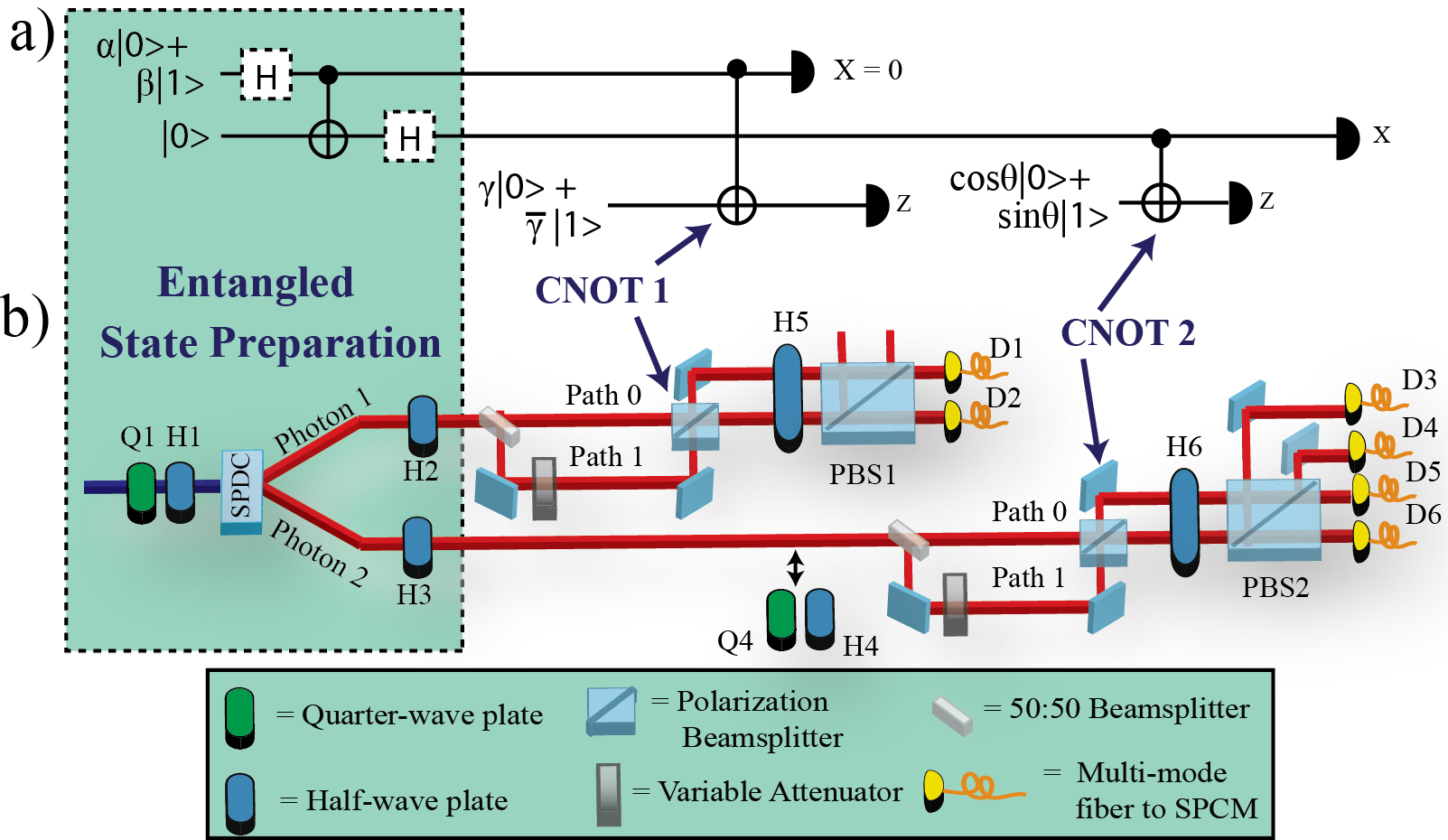}
\caption{Our implementation. a) The logical quantum circuit that we actually implement. We use ideas from cluster state quantum computing, namely single-qubit teleportation, to implement successive quantum gates. The first shaded area represents the creation of the entangled resource. After a 2-qubit cluster is created, the first qubit controls the controlled not gate used as our weak interaction. After this it is measured and its state is teleported to qubit 2. Qubit 2 then interacts with a second probe, which we use for our von Neumann measurement. b) The optical setup we use to implement the quantum circuit in (a). We use two entangled photons generated from spontaneous parametric down-conversion as the first two qubits of the circuit. Path qubits are added to each photon with 50/50 beam splitters. The polarization qubits are measured using polarizing beam splitters.}
\end{figure}

\begin{equation}
\Delta \hat{A}\Delta \hat{B}\geq \frac{1}{2}|\langle [\hat{A},\hat{B}]\rangle|.
\end{equation}
This form has been experimentally verified in many settings \cite{14}, and is uncontroversial. The corresponding generalization of Heisenberg's original measurement-disturbance relationship (MDR) would read:
\begin{equation}
\epsilon(\hat{A})\eta(\hat{B})\geq \frac{1}{2}|\langle [\hat{A},\hat{B}]\rangle|.
\end{equation}
This equation has been proven to be formally incorrect \cite{5}. Recently, Ozawa proved that the correct form of the MDR in fact reads \cite{6}:
\begin{equation}
\epsilon(\hat{A})\eta(\hat{B})  +\epsilon(\hat{A})\Delta\hat{B} + \eta(\hat{B})\Delta\hat{A} \geq \frac{1}{2}|\langle [\hat{A},\hat{B}]\rangle|.
\end{equation}
Due to the two additional terms on the left-hand side, this inequality may be satisfied even when Heisenberg's MDR is violated.

Experimentally observing a violation of Heisenberg's original MDR requires measuring the disturbance and precision of a measurement apparatus (MA). While classically measuring the disturbance is straightforward -- it simply requires knowing the value of an observable, $\hat{B}$, before and after the MA -- quantum mechanically it seems impossible. Quantum mechanics dictates that any attempt to measure $\hat{B}$ before the MA must disturb $\hat{B}$ (unless the system is already in an eigenstate of $\hat{B}$); as we shall discuss later, it may also change the state in such a way that the right-hand side (RHS) of Heisenberg's inequality is modified as well. Due to these difficulties the disturbance, as described here, has been claimed to be experimentally inaccessible \cite{15}. A recent experiment has indirectly tested Ozawa's new MDR \cite{7}, using a method also proposed by Ozawa \cite{16}. Rather than directly characterizing the effects of an individual measurement, this work checked the consistency of Ozawa's theory by carrying out a set of measurements from which the disturbance could be inferred through tomographic means \cite{17}; there has been some discussion on the arXiv as to the validity of this approach \cite{17,18,19,20}.   In contrast, Lund and Wiseman showed that if the system is weakly measured \cite{21,22} before the MA (figure 1a) the precision and disturbance can be directly observed in the resulting weak values \cite{23}. Here we present an experimental realization of this proposal, directly measuring the precision of an MA and its resulting disturbance, and demonstrate a clear violation of Heisenberg's MDR.

To understand the definitions of the precision and disturbance we first describe our implementation of a variable-strength measurement. A variable-strength measurement can be realized as an interaction between the system and a probe followed by a strong measurement of the probe \cite{24} (shaded area of figure 1a). The system and probe become entangled through the interaction, disturbing the system, such that measuring the probe will yield information about the state of system.  We define the disturbance as the root mean squared (RMS) difference between the value of $\hat{B}$ on the system before and after the MA, while the precision is the RMS difference between the value of $\hat{A}$ on the system before the interaction and the value of $\hat{A}$ read out on the probe. Lund and Wiseman showed these RMS differences can be directly obtained from a weak measurement on the system before the MA, post-selected on a projective measurement on either the probe or system afterwards \cite{23}. Specifically, they showed that the precision and disturbance for discrete variables is simply related to the weak-valued probabilities of $\hat{A}$ and $\hat{B}$ changing, $P_{WV}(\delta\hat{A})$ and $P_{WV}(\delta\hat{B})$, via:
\begin{eqnarray}
\epsilon(\hat{A})^2= \Sigma_{\delta\hat{A}}(\delta\hat{A})^2  P_{WV}(\delta\hat{A})\\
\eta(\hat{B})^2= \Sigma_{\delta\hat{B}}(\delta\hat{B})^2  P_{WV}(\delta\hat{B}).
\end{eqnarray}

By taking our system to be the polarization of a single photon we can demonstrate a violation of Heisenberg's precision limit by measuring one polarization component, $\hat{Z}$, and observing the resulting disturbance imparted to another, $\hat{X}$. Here, $\hat{X}$, $\hat{Y}$ and $\hat{Z}$  are the different polarization components of the photon; we use the convention that their eigenvalues are $\pm1$. For these observables, the bound (RHS of equations 2 and 3) of both Heisenberg and Ozawa's precision limits is $|\langle \hat{Y}\rangle|$. To facilitate the demonstration of a violation of Heisenberg's MDR, we make this bound as large as possible by preparing the system in the state $(\ket{H}+i\ket{V})/\sqrt{2}$, so that $|\langle \hat{Y}\rangle|=1$. In this state, the uncertainties are $\Delta\hat{X}=\Delta\hat{Z}=1$, which satisfy Heisenberg's uncertainty principle (equation 1), as they must. On the other hand, a measurement of  $\Delta\hat{Z}$ can be made arbitrarily precise. Now, even if the Z-precision, $\epsilon(\hat{Z})$, approaches zero the X-disturbance, $\eta(\hat{X})$, to $\hat{X}$ can only be as large as $\sqrt{2}$, so that their product can fall below 1, violating Heisenberg's MDR. Note that attempting the same violation with the Heisenberg uncertainty principle, by setting $\Delta\hat{Z}$ to zero, requires that the system is prepared in either $\ket{H}$ or $\ket{V}$, in which case the bound, $|\langle \hat{Y}\rangle|$, must also go to zero, so that equation 1 is trivially satisfied.

We can measure $\hat{Z}$ of a single photon, by coupling it to a probe system with a quantum logic gate \cite{25} (shaded region of figure 1b), implemented in additional path degrees of freedom of the photon \cite{26}. We use this technique to implement both the weak measurement and the MA. Current linear-optical quantum gates are reliant on post-selection, which makes them prone to error \cite{27}. We circumvent this problem, making use of ideas from the one-way model of quantum computing to implement the quantum circuit of figure 1b \cite{28}. To enable successive CNOT gates between the system and the two probes we first make a ``2-qubit line cluster'' in the polarization of two photons. 

Experimentally, we generate entangled 2-photon states of the form $\alpha\ket{HH}+\beta\ket{VV}$, using a spontaneous parametric down-conversion source in the ``sandwich-configuration'' \cite{29}. Each crystal is a 1mm crystal of BBO, cut for type-I phase matching. We can set $\alpha$ and $\beta$ by setting the pump polarization with quarter- and half-wave plates Q1 and H1 (figure 2b). The pump beam is centered at 404 nm, with a power of 500 mW, generating down-converted photons at 808 nm. The pump is generated by frequency doubling a femtosecond Ti:sapph laser, which is centered at 808 nm, using a 2 mm long crystal of BBO. The down-converted photons are coupled into single-mode fiber before being sent to the rest of the experiment. We observe approximately 15000 entangled pairs a second, with $12\%$ coupling efficiency, directly in the fiber. When coupling the light into multi-mode fiber after the interferometers, we measure about 1000 coincidence counts a second, spread among all the detector pairs. For each data point we acquire coincidence counts for 30 seconds using a homebuilt coincidence counter based on an FPGA. We are able to make the desired entangled state with a fidelity of $95.9\%$, which we measure by performing quantum state tomography (QST) on the photons directly after the single-mode fiber using a standard polarization tomography setup \cite{30}.

A modified quantum circuit which implements Lund and Wiseman's proposal \cite{23} and includes the line cluster creation is drawn in figure 2a, with the corresponding optical implementation below in figure 2b. A single logical polarisation qubit, $\alpha\ket{H}+\beta\ket{V}$, is encoded in two physical polarization qubits, forming the line cluster $\alpha\ket{H_1H_2}+\beta\ket{V_1V_2}$. Using a line cluster allows the first photon's polarization to control a CNOT gate with an additional path degree of freedom, realized using a polarizing beam splitter (PBS), to implement the weak measurement. After this step the state is $\alpha\ket{H_1H_2}\ket{A_1}+\beta\ket{V_1V_2}\ket{B_1}$, where $\ket{A_1}$ and $\ket{B_1}$ denote two different states of the path degrees of freedom,$\ket{A_1}=\gamma\ket{P_0}+\overline{\gamma}\ket{P_1}$ and $\ket{B_1}=\overline{\gamma}\ket{P_0}+\gamma\ket{P_1}$. Now, measuring the first polarization in the $\hat{X}$ basis and finding $\hat{X}=+1$ teleports the state of the system to the polarization of the second photon,  $\bra{\frac{H_1+V_1}{\sqrt{2}}}(\alpha\ket{H_1H_2}\ket{A_1}+\beta\ket{V_1V_2}\ket{B_1})=\alpha\ket{H_2}\ket{A_1}+\beta\ket{V_2}\ket{B_1}$. (If instead, the measurement result is $\hat{X}=-1$ the teleported state will be unitarily rotated to $\alpha\ket{H_2}\ket{A_1}-\beta\ket{V_2}\ket{B_1}$; in principle, one could correct this using feed-forward \cite{31}, but for simplicity we discard these events.) We characterize the teleportation by performing QST on the teleported single photon polarisation. To do this we insert quarter- and half-wave plates, Q4 and H4, and remove the path qubit of photon 2. We find the teleported state has a fidelity of $93.4\%$ with the expected state, mainly due to the reduced visibility of the interferometers. The polarization of the second photon is now free to be measured by the MA, which is implemented using a PBS and additional path degrees of freedom of photon 2, in the same way that photon 1 was weakly measured.
\begin{figure}
\includegraphics[scale=0.5]{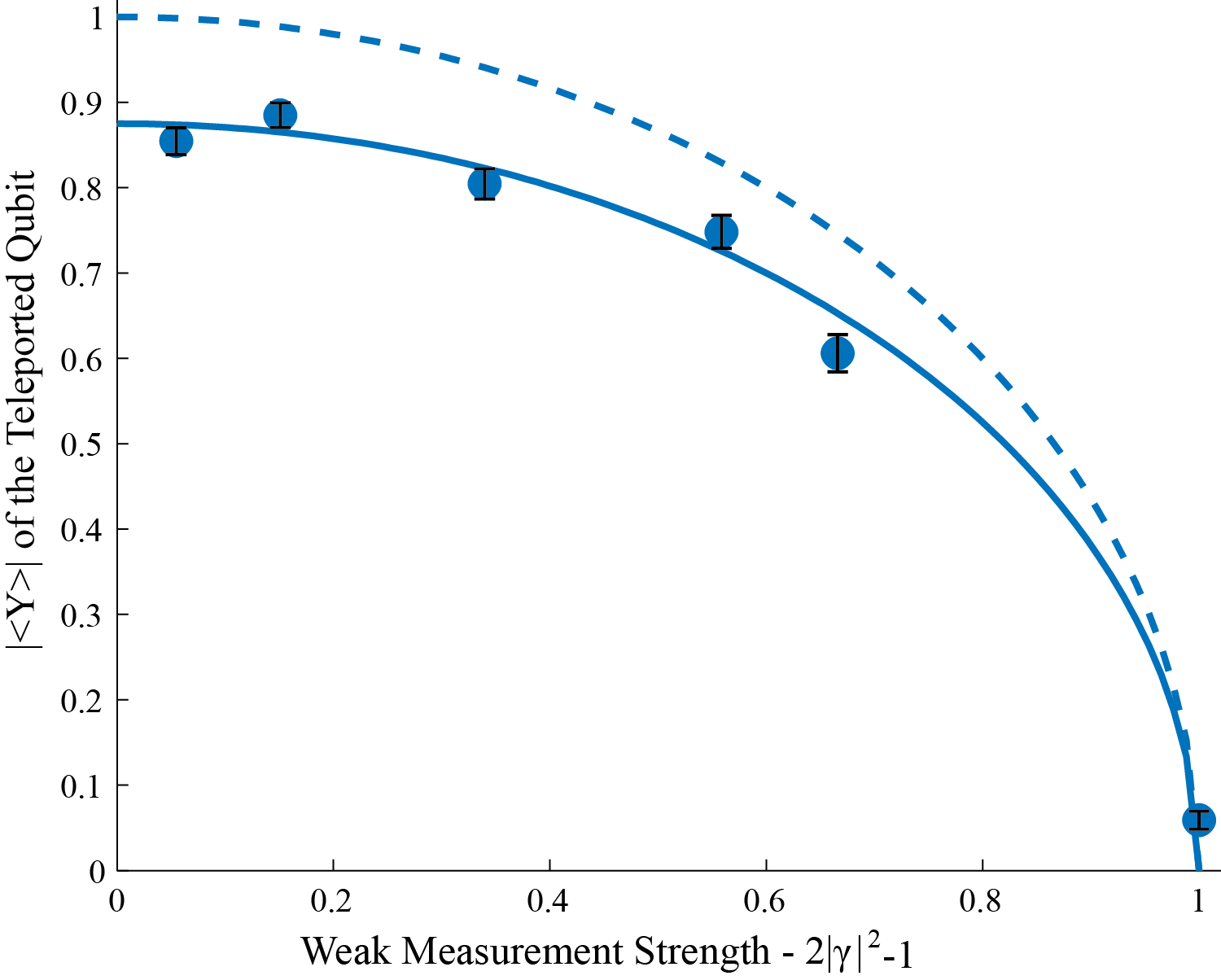}
\caption{Analysis of the bound. A plot of the RHS of equations 1 and 3 versus the strength of the weak probing measurement. The dashed line includes only the effect of the non-zero weak measurement strength. In addition to this effect, the solid line takes into account the imperfect teleportation, which agrees well with the experimentally measured points.}
\end{figure}

In order to clearly demonstrate a violation of Heisenberg's MDR we first experimentally characterize the bound of equations 2 and 3. Lund and Wiseman discuss the limiting case of using perfectly weak measurements to characterize the system before the action of the MA \cite{23}. However, in order to extract any information from this initial measurement, it cannot of course be infinitely weak. Although for our system, both the precision and the disturbance are independent of the weak measurement strength, the bound of equations 2 and 3 is not. For instance, if we replaced the weak measurement of $\hat{Z}$ with a strong one, this would project the system onto eigenstates of $\hat{Z}$, all of which have $|\langle \hat{Y}\rangle|=0$; the inequality would automatically be satisfied in this case.  The weaker the measurement, the less $|\langle \hat{Y}\rangle|$ is reduced, and the stronger the inequality.  We measured this experimentally, and figure 3 presents our data for $|\langle \hat{Y}\rangle|$ of the state just after the weak measurement, as a function of measurement strength, along with theory. It is important to note that these experimental difficulties can only lower the LHS of equation 2, and therefore cannot lead to a false violation.
\begin{figure}
\includegraphics[scale=0.55]{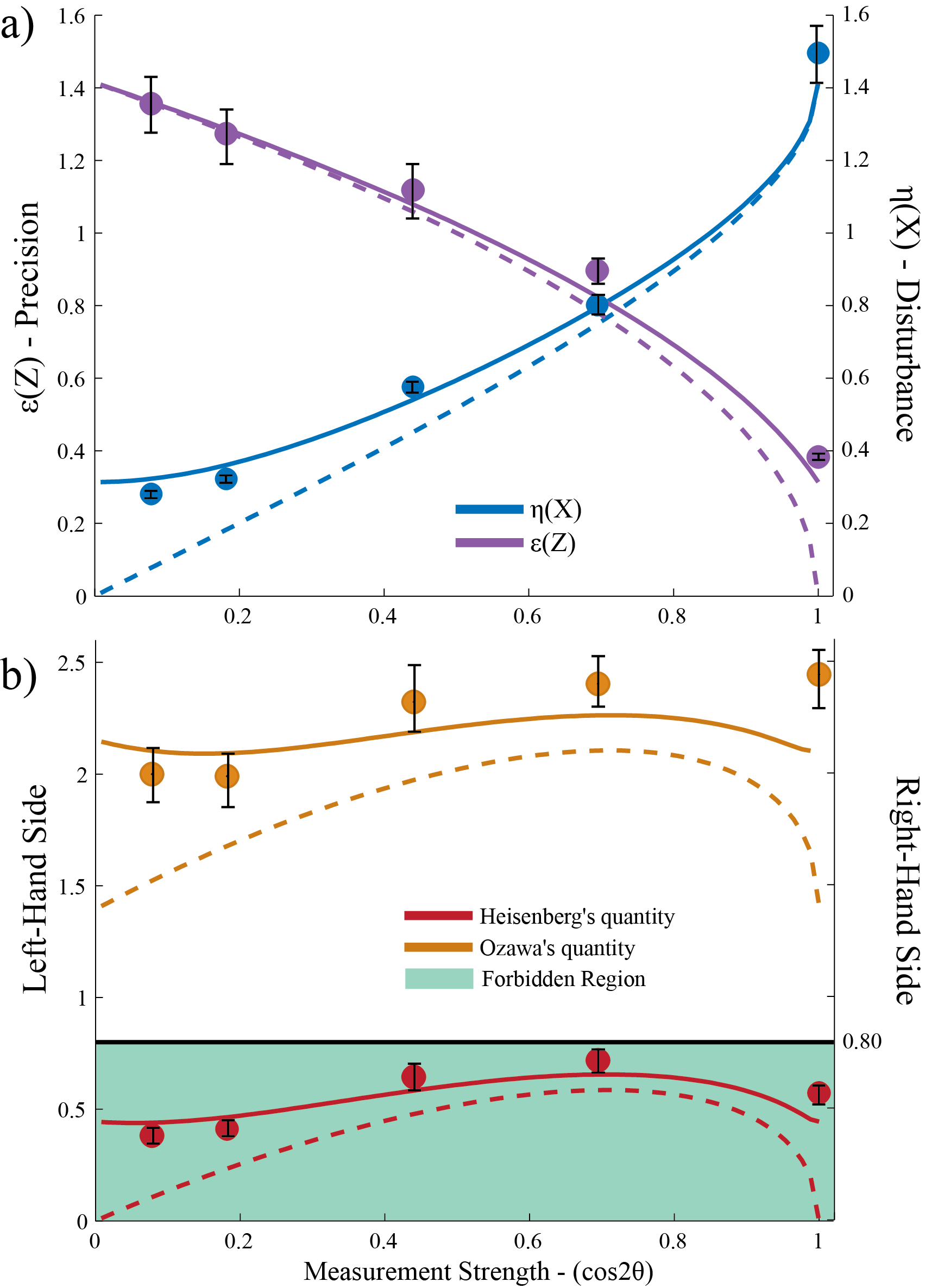}
\caption{Experimental results. a) The precision of the measurement apparatus (MA) and disturbance it imparts to the system plotted against its strength. b) A plot of the left-hand side of Heisenberg and Ozawa's relations versus the strength of the MA. For $\hat{X}$ and $\hat{Z}$ Heisenberg's quantity is $\epsilon(\hat{Z})\eta(\hat{X})$, and Ozawa's quantity is $\epsilon(\hat{Z})\eta(\hat{X})  +\epsilon(\hat{Z})\Delta\hat{X} + \eta(\hat{X})\Delta\hat{Z}$. The presumed bound on these quantities is given by the right-hand side of the relations, measured to be $|\langle \hat{Y}\rangle|=0.80\pm0.02$. Heisenberg's MDR is clearly violated, with his quantity falling below the bound, while Ozawa's MDR remains valid for all experimentally accessible parameters.}
\end{figure}

To show a violation of Heisenberg's MDR we measure the precision and the disturbance of the MA. To measure the X-disturbance we weakly measure $\hat{X}$  on the system before the MA post-selected on a strong measurement of $\hat{X}$ afterwards. Similarly, the Z-precision of the MA is obtained by weakly measuring $\hat{Z}$  and then post-selecting on a strong measurement of $\hat{Z}$ on the probe. From the results of these weak measurements the X-disturbance and Z-precision can be acquired. As an example, consider the X-disturbance, $\eta(\hat{X})$, as defined in equation 5. We need to measure the quantities $P_{WV}(\delta\hat{X})$ for all $\delta\hat{X}$. Since we are dealing with the polarisation of a single photon, $\delta\hat{X}$ can only equal 0 or $\pm2$. $P_{WV}(\delta\hat{X}=\pm2)$ is the weak probability that the system initially had $\hat{X}=\mp1$ and we found it in in $\hat{X}=\pm1$. These  probabilities can be expressed in terms of weak expectation values of $\hat{X}$, post-selected on finding the system after the MA with $\hat{X_f}=\pm1$, $\langle\hat{X}\rangle_{\hat{X_f}}$, as \cite{23}:
\begin{equation}
P_{WV}(\delta\hat{X}=\pm2) = \frac{1}{2}(1\mp\langle\hat{X}\rangle_{\hat{X_f}=\pm1})P(\hat{X_f}=\pm 1),
\end{equation}
In our experiment, $P(\hat{X_f}=+1)$ corresponds to the probability of finding photon 2 diagonally polarized, given that the teleportation on the first photon's polarization succeeds, which is signalled by photon 1 being diagonally polarized. As shown in figure 2b, both PBS's are set to measure in the diagonal/anti-diagonal basis, so this measurement amounts to counting two-photon events between the transmitted ports of PBS1 (detectors D1 or D2) and PBS 2 (detectors D5 or D6). The weak expectation value can be expressed in terms of the weak probe observable $\hat{Z_p}$, since $\hat{X}$ of the system couples to $\hat{Z}$ of the probe, as \cite{25}:
\begin{equation}
\langle\hat{X}\rangle_{\hat{X_f}}=\frac{P(\hat{Z_p}=+1|\hat{X_f})-P(\hat{Z_p}=-1|\hat{X_f})}{2|\gamma|^2-1}.
\end{equation}
Here, $2|\gamma|^2-1$ is the strength of the initial weak measurement, which we know and set through the state of the probe. The remaining quantities, $P(\hat{Z_p}=+1|\hat{X_f}=\pm1)$ and $P(\hat{Z_p}=-1|\hat{X_f}=\pm1)$, are directly measurable. For example, $P(\hat{Z_p}=+1|\hat{X_f}=\pm1)$ is the probability of finding the first photon in path 1 given that the second photon was found vertically polarized in either path. It is measured by two-photon events between detector D1 (for the teleportation to succeed and for $\hat{Z_p}=+1$) and the transmitted port of PBS 2 (detectors D5 or D6), to post-select on $\hat{X_f}=+1$. A similar analysis can be done for the Z-precision, but now rather than post-selecting on the polarization of photon 2, $\hat{X_f}$, one has to post-select on the $\hat{Z}$ value of the MA probe, which is the path of the second photon.

The precision and disturbance were measured for several measurement apparatus strengths and are plotted in figure 4a. The dashed lines are predictions for an ideal implementation of the quantum circuit in figure 3a, while the solid lines, which fit our data well, take into account the imperfect entangled state preparation. The imperfect state preparation leads to errors in the single-qubit teleportation, increasing the RMS difference between the measurements on the weak probe before the MA and the final verification measurements, on the system and probe, after the MA. Again, these errors can only increase disturbance and precision, and thus the LHS of equation 2, and cannot lead to a false violation.

From the measured precision and disturbance the LHS of Heisenberg and Ozawa's precision limits can be constructed. We set the strength of the initial weak measurement such that the RHS of equation 2 is large enough that Heisenberg's MDR violated for all settings of the MA. We measure $|\langle \hat{Y}\rangle|=0.80\pm0.02$, which gives the forbidden region in figure 4b. Heisenberg's quantity, which can be reconstructed simply from the measurements of the precision and the disturbance, is plotted in red. Ozawa's quantity, for which additional measurements of $\Delta\hat{X}$ and $\Delta\hat{Z}$ were made on the state, using quarter- and half-wave plates Q4 and H4, after the weak measurement, is plotted in orange. The error bars are due to Poissonian counting statistics. As seen in figure 4b, Ozawa's MDR remains valid for all the experimentally tested parameters, while we find that the simple product of the precision and the disturbance - Heisenberg's MDR- always falls below the experimentally measured bound. 

In conclusion, using weak measurements to experimentally characterize a system before and after it interacts with a measurement apparatus, we have directly measured its precision and the disturbance. This has allowed us to measure a violation of Heisenberg's hypothesized MDR. Our work conclusively shows that, although correct for uncertainties in states, the form of Heisenberg's precision limit is incorrect if naively applied to measurement. Our work highlights an important fundamental difference between uncertainties in states and the limitations of measurement in quantum mechanics.

We acknowledge financial support from the Natural Sciences and Engineering Research Council of Canada and the Candian Institute for Advanced Research.

\bibliography{rozema_bib}

\begin{thebibliography}{31}%
\makeatletter
\providecommand \@ifxundefined [1]{%
 \@ifx{#1\undefined}
}%
\providecommand \@ifnum [1]{%
 \ifnum #1\expandafter \@firstoftwo
 \else \expandafter \@secondoftwo
 \fi
}%
\providecommand \@ifx [1]{%
 \ifx #1\expandafter \@firstoftwo
 \else \expandafter \@secondoftwo
 \fi
}%
\providecommand \natexlab [1]{#1}%
\providecommand \enquote  [1]{``#1''}%
\providecommand \bibnamefont  [1]{#1}%
\providecommand \bibfnamefont [1]{#1}%
\providecommand \citenamefont [1]{#1}%
\providecommand \href@noop [0]{\@secondoftwo}%
\providecommand \href [0]{\begingroup \@sanitize@url \@href}%
\providecommand \@href[1]{\@@startlink{#1}\@@href}%
\providecommand \@@href[1]{\endgroup#1\@@endlink}%
\providecommand \@sanitize@url [0]{\catcode `\\12\catcode `\$12\catcode
  `\&12\catcode `\#12\catcode `\^12\catcode `\_12\catcode `\%12\relax}%
\providecommand \@@startlink[1]{}%
\providecommand \@@endlink[0]{}%
\providecommand \url  [0]{\begingroup\@sanitize@url \@url }%
\providecommand \@url [1]{\endgroup\@href {#1}{\urlprefix }}%
\providecommand \urlprefix  [0]{URL }%
\providecommand \Eprint [0]{\href }%
\providecommand \doibase [0]{http://dx.doi.org/}%
\providecommand \selectlanguage [0]{\@gobble}%
\providecommand \bibinfo  [0]{\@secondoftwo}%
\providecommand \bibfield  [0]{\@secondoftwo}%
\providecommand \translation [1]{[#1]}%
\providecommand \BibitemOpen [0]{}%
\providecommand \bibitemStop [0]{}%
\providecommand \bibitemNoStop [0]{.\EOS\space}%
\providecommand \EOS [0]{\spacefactor3000\relax}%
\providecommand \BibitemShut  [1]{\csname bibitem#1\endcsname}%
\let\auto@bib@innerbib\@empty
\bibitem [{\citenamefont {Heisenberg}(1927)}]{1}%
  \BibitemOpen
  \bibfield  {author} {\bibinfo {author} {\bibfnamefont {W.}~\bibnamefont
  {Heisenberg}},\ }\href@noop {} {\bibfield  {journal} {\bibinfo  {journal} {Z.
  Phys.}\ }\textbf {\bibinfo {volume} {43}},\ \bibinfo {pages} {172} (\bibinfo
  {year} {1927})},\ \bibinfo {note} {english translation in Quantum Theory and
  Measurement, J. A. Wheeler and W. H. Zurek, Eds., (Princeton Univ. Press,
  1984), pp. 62-84.}\BibitemShut {Stop}%
\bibitem [{\citenamefont {Robertson}(1929)}]{2}%
  \BibitemOpen
  \bibfield  {author} {\bibinfo {author} {\bibfnamefont {H.~P.}\ \bibnamefont
  {Robertson}},\ }\href@noop {} {\bibfield  {journal} {\bibinfo  {journal}
  {Phys. Rev.}\ }\textbf {\bibinfo {volume} {34}},\ \bibinfo {pages} {16}
  (\bibinfo {year} {1929})}\BibitemShut {NoStop}%
\bibitem [{\citenamefont {Kennard}(1927)}]{3}%
  \BibitemOpen
  \bibfield  {author} {\bibinfo {author} {\bibfnamefont {E.~H.}\ \bibnamefont
  {Kennard}},\ }\href@noop {} {\bibfield  {journal} {\bibinfo  {journal} {Z.
  Phys.}\ }\textbf {\bibinfo {volume} {44}},\ \bibinfo {pages} {326} (\bibinfo
  {year} {1927})}\BibitemShut {NoStop}%
\bibitem [{\citenamefont {Weyl}(1928)}]{4}%
  \BibitemOpen
  \bibfield  {author} {\bibinfo {author} {\bibfnamefont {H.}~\bibnamefont
  {Weyl}},\ }\href@noop {} {\emph {\bibinfo {title} {Gruppentheorie Und
  Quantenmechanik}}}\ (\bibinfo  {publisher} {Hirzel},\ \bibinfo {address}
  {Leipzig},\ \bibinfo {year} {1928})\BibitemShut {NoStop}%
\bibitem [{\citenamefont {Ballentine}(1970)}]{5}%
  \BibitemOpen
  \bibfield  {author} {\bibinfo {author} {\bibfnamefont {L.~E.}\ \bibnamefont
  {Ballentine}},\ }\href@noop {} {\bibfield  {journal} {\bibinfo  {journal}
  {Rev. Mod. Phys.}\ }\textbf {\bibinfo {volume} {42}},\ \bibinfo {pages} {358}
  (\bibinfo {year} {1970})}\BibitemShut {NoStop}%
\bibitem [{\citenamefont {Ozawa}(2003)}]{6}%
  \BibitemOpen
  \bibfield  {author} {\bibinfo {author} {\bibfnamefont {M.}~\bibnamefont
  {Ozawa}},\ }\href@noop {} {\bibfield  {journal} {\bibinfo  {journal} {Phys.
  Rev. A}\ }\textbf {\bibinfo {volume} {67}},\ \bibinfo {pages} {042105}
  (\bibinfo {year} {2003})}\BibitemShut {NoStop}%
\bibitem [{\citenamefont {Erhart}\ \emph
  {et~al.}(2012{\natexlab{a}})\citenamefont {Erhart}, \citenamefont {Sponar},
  \citenamefont {Sulyok}, \citenamefont {Badurek}, \citenamefont {Ozawa},\ and\
  \citenamefont {Hasegawa}}]{7}%
  \BibitemOpen
  \bibfield  {author} {\bibinfo {author} {\bibfnamefont {J.}~\bibnamefont
  {Erhart}}, \bibinfo {author} {\bibfnamefont {S.}~\bibnamefont {Sponar}},
  \bibinfo {author} {\bibfnamefont {G.}~\bibnamefont {Sulyok}}, \bibinfo
  {author} {\bibfnamefont {G.}~\bibnamefont {Badurek}}, \bibinfo {author}
  {\bibfnamefont {M.}~\bibnamefont {Ozawa}}, \ and\ \bibinfo {author}
  {\bibfnamefont {Y.}~\bibnamefont {Hasegawa}},\ }\href@noop {} {\bibfield
  {journal} {\bibinfo  {journal} {Nature Phys.}\ }\textbf {\bibinfo {volume}
  {8}},\ \bibinfo {pages} {185} (\bibinfo {year}
  {2012}{\natexlab{a}})}\BibitemShut {NoStop}%
\bibitem [{\citenamefont {Lund}\ and\ \citenamefont {Wiseman}(2010)}]{23}%
  \BibitemOpen
  \bibfield  {author} {\bibinfo {author} {\bibfnamefont {A.~P.}\ \bibnamefont
  {Lund}}\ and\ \bibinfo {author} {\bibfnamefont {H.~M.}\ \bibnamefont
  {Wiseman}},\ }\href@noop {} {\bibfield  {journal} {\bibinfo  {journal} {New
  J. Phys.}\ }\textbf {\bibinfo {volume} {12}},\ \bibinfo {pages} {093011}
  (\bibinfo {year} {2010})}\BibitemShut {NoStop}%
\bibitem [{\citenamefont {Wheeler}\ and\ \citenamefont {Zurek}(1984)}]{8}%
  \BibitemOpen
  \bibinfo {editor} {\bibfnamefont {J.~A.}\ \bibnamefont {Wheeler}}\ and\
  \bibinfo {editor} {\bibfnamefont {W.~H.}\ \bibnamefont {Zurek}},\ eds.,\
  \enquote {\bibinfo {title} {Quantum theory and measurement},}\ \ (\bibinfo
  {publisher} {Princeton Univ. Press},\ \bibinfo {year} {1984})\ pp.\ \bibinfo
  {pages} {3--49},\ \bibinfo {note} {(N. Bohr and A. Einstein)}\BibitemShut
  {NoStop}%
\bibitem [{\citenamefont {Scully}\ \emph {et~al.}(1991)\citenamefont {Scully},
  \citenamefont {Englert},\ and\ \citenamefont {Walther}}]{9}%
  \BibitemOpen
  \bibfield  {author} {\bibinfo {author} {\bibfnamefont {M.~O.}\ \bibnamefont
  {Scully}}, \bibinfo {author} {\bibfnamefont {B.~G.}\ \bibnamefont {Englert}},
  \ and\ \bibinfo {author} {\bibfnamefont {H.}~\bibnamefont {Walther}},\
  }\href@noop {} {\bibfield  {journal} {\bibinfo  {journal} {Nature}\ }\textbf
  {\bibinfo {volume} {351}},\ \bibinfo {pages} {111} (\bibinfo {year}
  {1991})}\BibitemShut {NoStop}%
\bibitem [{\citenamefont {Storey}\ \emph {et~al.}(1994)\citenamefont {Storey},
  \citenamefont {Tan}, \citenamefont {Collett},\ and\ \citenamefont
  {Walls}}]{10}%
  \BibitemOpen
  \bibfield  {author} {\bibinfo {author} {\bibfnamefont {P.}~\bibnamefont
  {Storey}}, \bibinfo {author} {\bibfnamefont {S.}~\bibnamefont {Tan}},
  \bibinfo {author} {\bibfnamefont {M.}~\bibnamefont {Collett}}, \ and\
  \bibinfo {author} {\bibfnamefont {D.}~\bibnamefont {Walls}},\ }\href@noop {}
  {\bibfield  {journal} {\bibinfo  {journal} {Nature}\ }\textbf {\bibinfo
  {volume} {367}},\ \bibinfo {pages} {626} (\bibinfo {year}
  {1994})}\BibitemShut {NoStop}%
\bibitem [{\citenamefont {Mir}\ \emph {et~al.}(2007)\citenamefont {Mir},
  \citenamefont {Lundeen}, \citenamefont {Mitchell}, \citenamefont {Steinberg},
  \citenamefont {Wiseman},\ and\ \citenamefont {Garretson}}]{11}%
  \BibitemOpen
  \bibfield  {author} {\bibinfo {author} {\bibfnamefont {R.}~\bibnamefont
  {Mir}}, \bibinfo {author} {\bibfnamefont {J.}~\bibnamefont {Lundeen}},
  \bibinfo {author} {\bibfnamefont {M.}~\bibnamefont {Mitchell}}, \bibinfo
  {author} {\bibfnamefont {A.}~\bibnamefont {Steinberg}}, \bibinfo {author}
  {\bibfnamefont {H.}~\bibnamefont {Wiseman}}, \ and\ \bibinfo {author}
  {\bibfnamefont {J.}~\bibnamefont {Garretson}},\ }\href@noop {} {\bibfield
  {journal} {\bibinfo  {journal} {New J. Phys.}\ }\textbf {\bibinfo {volume}
  {9}},\ \bibinfo {pages} {287} (\bibinfo {year} {2007})}\BibitemShut {NoStop}%
\bibitem [{\citenamefont {Prevedel}\ \emph {et~al.}(2011)\citenamefont
  {Prevedel}, \citenamefont {Hamel}, \citenamefont {Colbeck}, \citenamefont
  {Fisher},\ and\ \citenamefont {Resch}}]{12}%
  \BibitemOpen
  \bibfield  {author} {\bibinfo {author} {\bibfnamefont {R.}~\bibnamefont
  {Prevedel}}, \bibinfo {author} {\bibfnamefont {D.~R.}\ \bibnamefont {Hamel}},
  \bibinfo {author} {\bibfnamefont {R.}~\bibnamefont {Colbeck}}, \bibinfo
  {author} {\bibfnamefont {K.}~\bibnamefont {Fisher}}, \ and\ \bibinfo {author}
  {\bibfnamefont {K.~J.}\ \bibnamefont {Resch}},\ }\href@noop {} {\bibfield
  {journal} {\bibinfo  {journal} {Nature Phys.}\ }\textbf {\bibinfo {volume}
  {7}},\ \bibinfo {pages} {757} (\bibinfo {year} {2011})}\BibitemShut {NoStop}%
\bibitem [{\citenamefont {Berta}\ \emph {et~al.}(2010)\citenamefont {Berta},
  \citenamefont {Christandl}, \citenamefont {Colbeck}, \citenamefont {Renes},\
  and\ \citenamefont {Renner}}]{13}%
  \BibitemOpen
  \bibfield  {author} {\bibinfo {author} {\bibfnamefont {M.}~\bibnamefont
  {Berta}}, \bibinfo {author} {\bibfnamefont {M.}~\bibnamefont {Christandl}},
  \bibinfo {author} {\bibfnamefont {R.}~\bibnamefont {Colbeck}}, \bibinfo
  {author} {\bibfnamefont {J.}~\bibnamefont {Renes}}, \ and\ \bibinfo {author}
  {\bibfnamefont {R.}~\bibnamefont {Renner}},\ }\href@noop {} {\bibfield
  {journal} {\bibinfo  {journal} {Nature Phys.}\ }\textbf {\bibinfo {volume}
  {6}},\ \bibinfo {pages} {659} (\bibinfo {year} {2010})}\BibitemShut {NoStop}%
\bibitem [{\citenamefont {Nairz}\ \emph {et~al.}(2002)\citenamefont {Nairz},
  \citenamefont {Arndt},\ and\ \citenamefont {Zeilinger}}]{14}%
  \BibitemOpen
  \bibfield  {author} {\bibinfo {author} {\bibfnamefont {O.}~\bibnamefont
  {Nairz}}, \bibinfo {author} {\bibfnamefont {M.}~\bibnamefont {Arndt}}, \ and\
  \bibinfo {author} {\bibfnamefont {A.}~\bibnamefont {Zeilinger}},\ }\href@noop
  {} {\bibfield  {journal} {\bibinfo  {journal} {Phys. Rev. A}\ }\textbf
  {\bibinfo {volume} {65}},\ \bibinfo {pages} {032109} (\bibinfo {year}
  {2002})},\ \bibinfo {note} {and references therein.}\BibitemShut {Stop}%
\bibitem [{\citenamefont {Werner}(2004)}]{15}%
  \BibitemOpen
  \bibfield  {author} {\bibinfo {author} {\bibfnamefont {R.~F.}\ \bibnamefont
  {Werner}},\ }\href@noop {} {\bibfield  {journal} {\bibinfo  {journal}
  {Quantum Inf. Comput.}\ }\textbf {\bibinfo {volume} {4}},\ \bibinfo {pages}
  {546} (\bibinfo {year} {2004})}\BibitemShut {NoStop}%
\bibitem [{\citenamefont {Ozawa}(2004)}]{16}%
  \BibitemOpen
  \bibfield  {author} {\bibinfo {author} {\bibfnamefont {M.}~\bibnamefont
  {Ozawa}},\ }\href@noop {} {\bibfield  {journal} {\bibinfo  {journal} {Ann.
  Phys.}\ }\textbf {\bibinfo {volume} {311}},\ \bibinfo {pages} {350} (\bibinfo
  {year} {2004})}\BibitemShut {NoStop}%
\bibitem [{\citenamefont {Hofmann}(2012)}]{17}%
  \BibitemOpen
  \bibfield  {author} {\bibinfo {author} {\bibfnamefont {H.}~\bibnamefont
  {Hofmann}},\ }\href@noop {} {\bibfield  {journal} {\bibinfo  {journal}
  {pre-print}\ } (\bibinfo {year} {2012})},\ \bibinfo {note}
  {arXiv:1205.0073}\BibitemShut {NoStop}%
\bibitem [{\citenamefont {Kurihara}(2012{\natexlab{a}})}]{18}%
  \BibitemOpen
  \bibfield  {author} {\bibinfo {author} {\bibfnamefont {Y.}~\bibnamefont
  {Kurihara}},\ }\href@noop {} {\bibfield  {journal} {\bibinfo  {journal}
  {pre-print}\ } (\bibinfo {year} {2012}{\natexlab{a}})},\ \bibinfo {note}
  {arXiv:1201.5151}\BibitemShut {NoStop}%
\bibitem [{\citenamefont {Erhart}\ \emph
  {et~al.}(2012{\natexlab{b}})\citenamefont {Erhart}, \citenamefont {Sponar},
  \citenamefont {Sulyok}, \citenamefont {Badurek}, \citenamefont {Ozawa},\ and\
  \citenamefont {Hasegawa}}]{19}%
  \BibitemOpen
  \bibfield  {author} {\bibinfo {author} {\bibfnamefont {J.}~\bibnamefont
  {Erhart}}, \bibinfo {author} {\bibfnamefont {S.}~\bibnamefont {Sponar}},
  \bibinfo {author} {\bibfnamefont {G.}~\bibnamefont {Sulyok}}, \bibinfo
  {author} {\bibfnamefont {G.}~\bibnamefont {Badurek}}, \bibinfo {author}
  {\bibfnamefont {M.}~\bibnamefont {Ozawa}}, \ and\ \bibinfo {author}
  {\bibfnamefont {Y.}~\bibnamefont {Hasegawa}},\ }\href@noop {} {\bibfield
  {journal} {\bibinfo  {journal} {pre-print}\ } (\bibinfo {year}
  {2012}{\natexlab{b}})},\ \bibinfo {note} {arXiv:1203.0927}\BibitemShut
  {NoStop}%
\bibitem [{\citenamefont {Kurihara}(2012{\natexlab{b}})}]{20}%
  \BibitemOpen
  \bibfield  {author} {\bibinfo {author} {\bibfnamefont {Y.}~\bibnamefont
  {Kurihara}},\ }\href@noop {} {\bibfield  {journal} {\bibinfo  {journal}
  {pre-print}\ } (\bibinfo {year} {2012}{\natexlab{b}})},\ \bibinfo {note}
  {arXiv: 1206.0421}\BibitemShut {NoStop}%
\bibitem [{\citenamefont {Aharonov}\ \emph {et~al.}(1988)\citenamefont
  {Aharonov}, \citenamefont {Albert},\ and\ \citenamefont {Vaidman}}]{21}%
  \BibitemOpen
  \bibfield  {author} {\bibinfo {author} {\bibfnamefont {Y.}~\bibnamefont
  {Aharonov}}, \bibinfo {author} {\bibfnamefont {D.~Z.}\ \bibnamefont
  {Albert}}, \ and\ \bibinfo {author} {\bibfnamefont {L.}~\bibnamefont
  {Vaidman}},\ }\href@noop {} {\bibfield  {journal} {\bibinfo  {journal} {Phys.
  Rev. Lett.}\ }\textbf {\bibinfo {volume} {60}},\ \bibinfo {pages} {1351}
  (\bibinfo {year} {1988})}\BibitemShut {NoStop}%
\bibitem [{\citenamefont {Kocsis}\ \emph {et~al.}(2011)\citenamefont {Kocsis},
  \citenamefont {Braverman}, \citenamefont {Ravets}, \citenamefont {Stevens},
  \citenamefont {Mirin}, \citenamefont {Shalm},\ and\ \citenamefont
  {Steinberg}}]{22}%
  \BibitemOpen
  \bibfield  {author} {\bibinfo {author} {\bibfnamefont {S.}~\bibnamefont
  {Kocsis}}, \bibinfo {author} {\bibfnamefont {B.}~\bibnamefont {Braverman}},
  \bibinfo {author} {\bibfnamefont {S.}~\bibnamefont {Ravets}}, \bibinfo
  {author} {\bibfnamefont {M.}~\bibnamefont {Stevens}}, \bibinfo {author}
  {\bibfnamefont {R.}~\bibnamefont {Mirin}}, \bibinfo {author} {\bibfnamefont
  {L.}~\bibnamefont {Shalm}}, \ and\ \bibinfo {author} {\bibfnamefont
  {A.}~\bibnamefont {Steinberg}},\ }\href@noop {} {\bibfield  {journal}
  {\bibinfo  {journal} {Science}\ }\textbf {\bibinfo {volume} {332}},\ \bibinfo
  {pages} {1170} (\bibinfo {year} {2011})}\BibitemShut {NoStop}%
\bibitem [{\citenamefont {von Neumann}(1955)}]{24}%
  \BibitemOpen
  \bibfield  {author} {\bibinfo {author} {\bibfnamefont {J.}~\bibnamefont {von
  Neumann}},\ }\href@noop {} {\emph {\bibinfo {title} {Mathematical Foundations
  of Quantum Mechanics}}}\ (\bibinfo  {publisher} {Princeton Univ. Press},\
  \bibinfo {year} {1955})\ \bibinfo {note} {originally published in 1932 as
  'Mathematische Grundlagen der Quantenmechanik'.}\BibitemShut {Stop}%
\bibitem [{\citenamefont {Pryde}\ \emph {et~al.}(2005)\citenamefont {Pryde},
  \citenamefont {O'Brien}, \citenamefont {White}, \citenamefont {Ralph},\ and\
  \citenamefont {Wiseman}}]{25}%
  \BibitemOpen
  \bibfield  {author} {\bibinfo {author} {\bibfnamefont {G.~J.}\ \bibnamefont
  {Pryde}}, \bibinfo {author} {\bibfnamefont {J.~L.}\ \bibnamefont {O'Brien}},
  \bibinfo {author} {\bibfnamefont {A.~G.}\ \bibnamefont {White}}, \bibinfo
  {author} {\bibfnamefont {T.~C.}\ \bibnamefont {Ralph}}, \ and\ \bibinfo
  {author} {\bibfnamefont {H.~M.}\ \bibnamefont {Wiseman}},\ }\href@noop {}
  {\bibfield  {journal} {\bibinfo  {journal} {Phys. Rev. Lett.}\ }\textbf
  {\bibinfo {volume} {94}},\ \bibinfo {pages} {220405} (\bibinfo {year}
  {2005})}\BibitemShut {NoStop}%
\bibitem [{\citenamefont {Iinuma}\ \emph {et~al.}(2011)\citenamefont {Iinuma},
  \citenamefont {Suzuki}, \citenamefont {Taguchi}, \citenamefont {Kadoya},\
  and\ \citenamefont {Hofmann}}]{26}%
  \BibitemOpen
  \bibfield  {author} {\bibinfo {author} {\bibfnamefont {M.}~\bibnamefont
  {Iinuma}}, \bibinfo {author} {\bibfnamefont {Y.}~\bibnamefont {Suzuki}},
  \bibinfo {author} {\bibfnamefont {G.}~\bibnamefont {Taguchi}}, \bibinfo
  {author} {\bibfnamefont {Y.}~\bibnamefont {Kadoya}}, \ and\ \bibinfo {author}
  {\bibfnamefont {H.~F.}\ \bibnamefont {Hofmann}},\ }\href@noop {} {\bibfield
  {journal} {\bibinfo  {journal} {New J. Phys.}\ }\textbf {\bibinfo {volume}
  {13}},\ \bibinfo {pages} {033041} (\bibinfo {year} {2011})}\BibitemShut
  {NoStop}%
\bibitem [{\citenamefont {O'Brien}\ \emph {et~al.}(2003)\citenamefont
  {O'Brien}, \citenamefont {Pryde}, \citenamefont {White}, \citenamefont
  {Ralph},\ and\ \citenamefont {Branning}}]{27}%
  \BibitemOpen
  \bibfield  {author} {\bibinfo {author} {\bibfnamefont {J.~L.}\ \bibnamefont
  {O'Brien}}, \bibinfo {author} {\bibfnamefont {G.~J.}\ \bibnamefont {Pryde}},
  \bibinfo {author} {\bibfnamefont {A.~G.}\ \bibnamefont {White}}, \bibinfo
  {author} {\bibfnamefont {T.~C.}\ \bibnamefont {Ralph}}, \ and\ \bibinfo
  {author} {\bibfnamefont {D.}~\bibnamefont {Branning}},\ }\href@noop {}
  {\bibfield  {journal} {\bibinfo  {journal} {Nature}\ }\textbf {\bibinfo
  {volume} {426}},\ \bibinfo {pages} {264} (\bibinfo {year}
  {2003})}\BibitemShut {NoStop}%
\bibitem [{\citenamefont {Raussendorf}\ and\ \citenamefont
  {Briegel}(2001)}]{28}%
  \BibitemOpen
  \bibfield  {author} {\bibinfo {author} {\bibfnamefont {R.}~\bibnamefont
  {Raussendorf}}\ and\ \bibinfo {author} {\bibfnamefont {H.}~\bibnamefont
  {Briegel}},\ }\href@noop {} {\bibfield  {journal} {\bibinfo  {journal} {Phys.
  Rev. Lett.}\ }\textbf {\bibinfo {volume} {86}},\ \bibinfo {pages} {5188}
  (\bibinfo {year} {2001})}\BibitemShut {NoStop}%
\bibitem [{\citenamefont {Kwiat}\ \emph {et~al.}(1999)\citenamefont {Kwiat},
  \citenamefont {Waks}, \citenamefont {White}, \citenamefont {Appelbaum},\ and\
  \citenamefont {Eberhard}}]{29}%
  \BibitemOpen
  \bibfield  {author} {\bibinfo {author} {\bibfnamefont {P.~G.}\ \bibnamefont
  {Kwiat}}, \bibinfo {author} {\bibfnamefont {E.}~\bibnamefont {Waks}},
  \bibinfo {author} {\bibfnamefont {A.~G.}\ \bibnamefont {White}}, \bibinfo
  {author} {\bibfnamefont {I.}~\bibnamefont {Appelbaum}}, \ and\ \bibinfo
  {author} {\bibfnamefont {P.~H.}\ \bibnamefont {Eberhard}},\ }\href@noop {}
  {\bibfield  {journal} {\bibinfo  {journal} {Phys. Rev. A.}\ }\textbf
  {\bibinfo {volume} {60}},\ \bibinfo {pages} {R773} (\bibinfo {year}
  {1999})}\BibitemShut {NoStop}%
\bibitem [{\citenamefont {James}\ \emph {et~al.}(2001)\citenamefont {James},
  \citenamefont {Kwiat}, \citenamefont {Munro},\ and\ \citenamefont
  {White}}]{30}%
  \BibitemOpen
  \bibfield  {author} {\bibinfo {author} {\bibfnamefont {D.~F.~V.}\
  \bibnamefont {James}}, \bibinfo {author} {\bibfnamefont {P.~G.}\ \bibnamefont
  {Kwiat}}, \bibinfo {author} {\bibfnamefont {W.~J.}\ \bibnamefont {Munro}}, \
  and\ \bibinfo {author} {\bibfnamefont {A.~G.}\ \bibnamefont {White}},\
  }\href@noop {} {\bibfield  {journal} {\bibinfo  {journal} {Phys. Rev. A.}\
  }\textbf {\bibinfo {volume} {64}},\ \bibinfo {pages} {052312} (\bibinfo
  {year} {2001})}\BibitemShut {NoStop}%
\bibitem [{\citenamefont {Prevedel}\ \emph {et~al.}(2007)\citenamefont
  {Prevedel}, \citenamefont {Walther}, \citenamefont {Tiefenbacher},
  \citenamefont {Böhi}, \citenamefont {Kaltenbaek}, \citenamefont {Jennewein},\
  and\ \citenamefont {Zeilinger}}]{31}%
  \BibitemOpen
  \bibfield  {author} {\bibinfo {author} {\bibfnamefont {R.}~\bibnamefont
  {Prevedel}}, \bibinfo {author} {\bibfnamefont {P.}~\bibnamefont {Walther}},
  \bibinfo {author} {\bibfnamefont {F.}~\bibnamefont {Tiefenbacher}}, \bibinfo
  {author} {\bibfnamefont {P.}~\bibnamefont {Böhi}}, \bibinfo {author}
  {\bibfnamefont {R.}~\bibnamefont {Kaltenbaek}}, \bibinfo {author}
  {\bibfnamefont {T.}~\bibnamefont {Jennewein}}, \ and\ \bibinfo {author}
  {\bibfnamefont {A.}~\bibnamefont {Zeilinger}},\ }\href@noop {} {\bibfield
  {journal} {\bibinfo  {journal} {Nature}\ }\textbf {\bibinfo {volume} {445}},\
  \bibinfo {pages} {65} (\bibinfo {year} {2007})}\BibitemShut {NoStop}%
\end{thebibliography}%

\end{document}